\documentclass[preprint,3p]{elsarticle}

\usepackage{amssymb}
\usepackage{amsmath}
\usepackage{amsbsy}
\usepackage{slashed}

\newcommand{\fort}  {\texttt{FORTRAN} }

\newcommand{\sarah} {{\tt SARAH} }

\newcommand{\spheno} {{\tt SPheno}}
\newcommand{\leer} [1] {}

\begin{document}

\title{
       {\tt \bf SPheno} 3.1: extensions including flavour, CP-phases
       and models beyond the MSSM}
\author{W.~Porod\footnote{porod@physik.uni-wuerzbzrg.de}
 and F.~Staub\footnote{fnstaub@physik.uni-wuerzbzrg.de} \\[0.4cm]
Institut f\"ur Theoretische Physik und Astrophysik,
Universit\"at W\"urzburg\\ D-97074 W\"urzburg, Germany}
\begin{abstract}
We describe  recent extensions of the program
\spheno\ including flavour aspects,
CP-phases, R-parity violation and low energy observables. 
In case of flavour mixing all
masses of supersymmetric particles are calculated including the complete
flavour structure and all possible CP-phases at the 1-loop level.
We give details on implemented
seesaw models, low energy observables and the corresponding extension
of the SUSY Les Houches Accord. Moreover, we comment on the
possiblities to include MSSM extensions in \spheno.
\end{abstract}

\maketitle


\section{Program Summary}
\noindent
{\bf Program title}: SPheno \\
{\bf Program Obtainable from:}
\verb+http://projects.hepforge.org/spheno/+\\
{\bf Programming language}: F95 \\
{\bf Computers for which the program has been designed}: 
PC running under Linux, should run in every Unix environment \\
{\bf Operating systems}: Linux, Unix \\
{\bf Keywords}: Supersymmetry,  renormalization group equations,
  mass spectra of supersymmetric models, Runge-Kutta, decays of supersymmetric particles, 
  production  \\
{\bf Nature of problem}:
 The first issue is the determination of the masses and couplings
 of supersymmetric particles in various supersymmetric models
 the R-parity conserved MSSM with generation mixing and including
 CP-violating phases, various seesaw extensions of the MSSM
 and the MSSM with bilinear R-parity breaking. Low energy
 data on Standard Model fermion masses, gauge couplings and electroweak
 gauge boson masses serve as constraints. Radiative corrections from
 supersymmetric particles to these inputs must be calculated. Theoretical
 constraints on the soft SUSY breaking parameters from a high scale
 theory are imposed and the parameters at the electroweak scale are obtained
 from the high scale parameters by evaluating the corresponding 
 renormalization group equations. These parameters must be consistent 
 with the requirement of correct electroweak symmetry breaking. 
 The second issue is to use the obtained masses and couplings for calculating
 decay widths and branching ratios of supersymmetric particles as well
 as the cross sections for theses particles in electron positron annihilation.
 The third issue is to calculate low energy constraints in the
 B-meson sector such as BR($b\to s \gamma)$, $\Delta M_{B_s}$,
 rare lepton decays,  such as BR($\mu\to e \gamma)$, the SUSY
 contributions to anomalous 
 magnetic moments and electric dipole moments of leptons,
  the SUSY contributions to the
 rho parameter as well as lepton flavour violating $Z$ decays.
 \\
{\bf Solution method}:  The renormalization connecting a high scale and the
 electroweak scale is calculated by the Runge-Kutta method. Iteration
 provides a solution consistent with with the multi-boundary conditions.
 In case of three-body decays and for the calculation of initial state
 radiation Gaussian quadrature is used for the numerical solution of the
 integrals. \\
 {\bf Restrictions:} In case of R-parity violation the cross sections
 are not calculated.\\
{\bf Running time}: 0.2 second on a Intel(R) Core(TM)2 Duo CPU T9900  with
 3.06GHz

\section{Introduction}

In its orginal version \spheno\ had been designed to calculate
the spectrum in the MSSM neglecting any effects due to
generation mixing and/or CP violation \cite{Porod:2003um}.
Moreover, the two- and three-body decays of the SUSY particles
as well as of the Higgs bosons can be calculated as well as the
production rates of these particles in $e^+ e^-$ annihilation.
The code itself is written in {\fort 95}.

The program has been extended to include
 flavour aspects, CP-violation and
R-parity violation. Moreover, different
variants of the seesaw mechanism
have been implemented. In this paper we describe the corresponding
changes and implementations.
Details on the algorithms used
can be found in the orginal manual \cite{Porod:2003um}.
Moreover, we give in the appendices the default values for various
flags as well as the error coding.

\section{Extensions with  MSSM particle content at the
electroweak scale and conserved R-parity}

\spheno\ has been extended  to include flavour and CP violating
phases using the SLHA2 conventions
\cite{Allanach:2008qq} for the general MSSM. 
For this purpose the complete flavour structures including
CP-phases have been implemented in the RGEs at the 2-loop level
using the formulas given in \cite{Martin:1993zk}.
We have extended
the formulas of \cite{Pierce:1996zz} for the 1-loop masses to account
for the flavour structures, e.g.\ we calculate the 1-loop corrected
$6\times 6$ mass matrices for squarks and charged leptons and the
1-loop corrected $3\times 3$ mass matrix for sneutrinos
\cite{Bruhnke:2010rh,Staub:2010ty}. Moreover,
we take into account all possible phases in the calculation of all
mass matrices at the 1-loop level but for one exception:
we do not consider the loop induced mixing between the scalar
and pseudoscalar Higgs bosons.

Beside the calculation of the spectrum also the decay routines have
been extended to calculate all possible two- and three body decays
of supersymmetric particles and Higgs bosons at leading order including
flavour effects. In addition the set of low energy observables has
been extended as described in section \ref{sec:lowenergy}.

Within the MSSM several model classes are implemented
\begin{itemize}
\item High scale models like mSUGRA, GMSB, AMSB 
\item A SUGRA scenario where all  soft SUSY breaking parameters
 are given freely at the GUT scale which is determined usually via
 the condition $g_1(M_{\rm GUT}) = g_2(M_{\rm GUT})$. However, this scale
 can be set to a fixed value using entry 31 in block 
 {\tt SPhenoInput}, see section \ref{sec:sphenoinput}.
\item All MSSM parameters specified at the electroweak scale $Q_{EWSB}$
with a user specified value for  $Q_{EWSB}$.
\end{itemize}
In all cases it is assumed that the required input is given
via the SLHA convention \cite{Skands:2003cj,Allanach:2008qq}.

In addition several
classes of neutrino mass models have been included with additional
states at  high energy scales:
\begin{itemize}
\item A seesaw I model with  different masses for the right-handed neutrinos.
  The corresponding particle content can be chosen setting the entry
  3 of the block {\tt MODSEL} as described in section \ref{sec:modsel}.
  The parameters are set using the blocks {\tt MNURIN} and {\tt YNU0IN},
  see sections \ref{sec:mnur} and \ref{sec:Ynu0}, respectively.
  Using this the results of \cite{Hirsch:2008dy,Esteves:2009vg} 
  have been obtained.
\item Two variants of  seesaw II model have been implemented.
Here one can either choose a pair of $SU(2)$ triplets or a pair
of $SU(5)$ 15-plets to generate neutrino masses.
The first  version  uses
the formulas  \cite{Rossi:2002zb} including
the corrections presented in \cite{Borzumati:2009hu} and
2-loop RGEs for the gauge couplings and gaugino mass parameters
as 
used in \cite{Hirsch:2008gh}. 
Here the blocks  {\tt Higgs3IN} and  {\tt YT0IN} 
have to be used to transfer the data,
see sections 
\ref{sec:blockhiggs3} and \ref{sec:blockyt0}. This variant is faster if
one uses 2-loop RGEs but less accurate in particular for low seesaw scales
\cite{Esteves:2010ff}.

In case of  a pair
of $SU(5)$ 15-plets a second variant has been implemented
using the complete 2-loop RGEs and corresponding threshold
corrections at the seesaw scale as described in \cite{Esteves:2009qr}.
Here the blocks  {\tt Higgs3IN} and  {\tt YTIN} 
have to be used to transfer the data,
see sections 
\ref{sec:blockhiggs3} and \ref{sec:blockytin}.

\item A seesaw III model with three $SU(5)$ matter
24-plets using the complete 2-loop RGEs and corresponding threshold
corrections at the seesaw scales as described in \cite{Esteves:2010ff}.
The blocks  {\tt MWMIN} and {\tt YB3IN}
 can be used to set the parameters, see sections
 \ref{sec:blockmwmin} and  \ref{sec:blockybin}.

\item A minimal $SU(5)$ model as described  in \cite{Deppisch:2007xu}.
  The corresponding particle content can be chosen setting the entry
  3 of the block {\tt MODSEL} as described in section \ref{sec:modsel}.
  The parameters are set using the blocks {\tt MNURIN} and {\tt YNU0IN},
  see sections \ref{sec:mnur} and \ref{sec:Ynu0}, respectively.
  The additional $SU(5)$ parameters can be set extending the
  block {\tt MINPAR} as described in section \ref{sec:minpar}.
\end{itemize}
 Note, that
in these models the particle content at the electroweak scale is
the same as in the usual MSSM and that the differences are only due to
the modified evaluation of the parameters.

\section{R-parity violation}

Currently  the bilinear model is implemented, i.e.\  extending
the superpotential by the terms
\begin{equation}
W_{\slashed{R}} = \epsilon_i \widehat{L}_i \widehat{H}_u
\end{equation}
and the corresponding soft SUSY breaking terms. In this class
of models neutrino physics can be explaining due to the mixing
of neutralinos with neutrinos and by loop contributions. The
corresponding details can be found in \cite{Hirsch:2000ef,Diaz:2003as}.
The same parameters giving rise to neutrino masses also lead
to the decay of the LSP and, thus, there are correlations between
the LSP decay properties and neutrino physics
\cite{Porod:2000hv,Hirsch:2002ys,Hirsch:2003fe,AristizabalSierra:2004cy,%
Hirsch:2005ag,deCampos:2007bn}.

Here one has two options
\begin{itemize}
\item use within SLHA2 the blocks {\tt EXTPAR}, {\tt RVSNVEVIN}
 and {\tt RVKAPPAIN}
   to specify the model parameters at the electroweak scale
\item use one of the high scale models mSUGRA, GMSB or AMSB to calculate
   the R-parity conserving parameters at the electroweak scale. The R-parity
   parameters are then added at this scale using one of the two possiblities
 \begin{itemize}
  \item add them using the blocks   {\tt RVSNVEVIN} and
   {\tt RVKAPPAIN}
  \item use the flag 91 of the block {\tt SPhenoInput} as
  described in section \ref{sec:sphenoinput} 
  to calculate the $\epsilon_i$ and the sneutrino vacuum expectations values $v_i$ such, that
    neutrino physics is respected. The corresponding neutrino data can 
    be specified in block {\tt NeutrinoBoundsIn},
    see section \ref{sec:neutrinoboundsin}.
 \end{itemize}
\end{itemize}
In this class of models the mass matrices are calculated at
tree-level except for the neutrino/neutralino mass matrix which
requires the inclusion of the full 1-loop contributions.
Moreover, all possible R-parity violating decay modes are
calculated. 

\section{Low energy observables}
\label{sec:lowenergy}

In this section we summarize the main references from which 
the formulas for the corresponding implementation have been
taken. Moreover we give implementation specific details whenever necessary.
For the moment  the low energy observables are only calculated
if there is an effective MSSM at the electroweak scale and
conserved R-parity. 

\subsection{B-physics observables}

The following observables are calculated in \spheno:
$BR(b \to s \gamma)$, $BR(b \to s \mu^+ \mu^-)$,
$BR(b \to s \sum_i \nu_i \nu_i)$,
 $BR(B^0_d \to \mu^+ \mu^-)$,
 $BR(B^0_s \to \mu^+ \mu^-)$,
 $BR(B_u \to \tau^+ \nu)$,
$\Delta M_{B_s^0}$  and $\Delta M_{B_d^0}$. For the calculation
of the Wilson coefficients we use running couplings and SUSY masses
which are in general evolved at the scale $Q=m_Z$. The only exception
is $BR(b \to s \gamma)$ as we use here the formula of ref.\
\cite{Lunghi:2006hc} where the corresponding coefficients have to be
given at the scale $Q=160$~GeV. For the calculation of the
Wilson coefficients and the corresponding observables we have used 
\begin{itemize}
\item 
$BR(b \to s \gamma)$ \cite{Bobeth:2001jm,Baek:2001kh,Lunghi:2006hc};
 the value given is for $E_\gamma \ge 1.6$~GeV and 
$m_c/m_b = 0.23$.
\item $BR(b \to s \mu^+ \mu^-)$ \cite{Bobeth:2001jm,Baek:2001kh,Huber:2005ig}
\item $BR(b \to s \sum_i \nu_i \nu_i)$ \cite{Baek:2001kh,Bobeth:2001jm}
\item  $BR(B^0_s \to \mu^+ \mu^-)$, $BR(B^0_d \to \mu^+ \mu^-)$
 \cite{Logan:2000iv,Baek:2001kh,Buras:2002vd}
\item  $BR(B_u \to \tau^+ \nu)$ \cite{Hou:1992sy}
\item $\Delta M_{B_s^0}$  and $\Delta M_{B_d^0}$
 \cite{Baek:2001kh,Buras:2002vd}. For the hadronic parameters we follow
\cite{Buras:2002vd}: 
\begin{equation}
\bar{P}^{LR}_1 = -0.71 \,\,,\,\,
\bar{P}^{LR}_2 = -0.9 \,\,,\,\,
\bar{P}^{SLL}_1 = -0.37 \,\,,\,\,
\bar{P}^{SLL}_1 = -0.72 \,\,.
\nonumber
\end{equation}

\end{itemize}
The remaining parameters used are given in table \ref{tab:hadronic}.
\begin{table}[t]
\caption{Parameters used in the calculation of the $B$-physics
observables.}
\label{tab:hadronic}
\begin{center}
\begin{tabular}{cccc}\hline
$\tau_{B^0}= 1.525$ ps & $\tau_{B^0_s}= 1.472$ ps & 
$\tau_{B^+_u}= 1.638$ ps & $\tau_{B^+_c}= 0.45$ ps  \\
$f_B = 190$~MeV & $B_{B_d}=1.22$ & $f_{B_s}=230$~MeV & $B_{B_s}=1.26$\\
$M_{B^0} = 5.2795$~GeV & $M_{B^0_s} = 5.3663$~GeV & $\eta_B=0.55$ \\
\hline
\end{tabular}
\end{center}
\end{table}

\subsection{Lepton sector}

In the leptonic sector a similar strategy is used: all parameters
are evolved to $m_Z$ and then running masses and mixing matrices 
are used as input for the observables. The implemented formulas
are based on
\begin{itemize}
\item SUSY contributions to the anomalous magnetic moment of the leptons
\cite{Ibrahim:1999hh}
\item electric dipole moments (EDMs) of the leptons \cite{Bartl:1999bc,Bartl:2003ju}
\item two body decays $\mu \to e \gamma$, $\tau \to e \gamma$
and $\tau \to \mu \gamma$
\cite{Hisano:1995cp,Bartl:2003ju}
\item three body decays $\mu \to e e^+ e^-$, $\tau \to e e^+ e^-$
and  $\tau \to \mu \mu^+ \mu^-$ \cite{Arganda:2005ji}
\item $Z$ decays, $Z \to e^\pm \mu^\mp$, $Z \to e^\pm \tau^\mp$ and
$Z \to \mu^\pm \tau^\mp$,  \cite{Bi:2000xp}
\end{itemize}

\subsection{Other constraints}

In addition the EDM of the neutron can be calculated using
two different models for the neutron where the formulas
are based on  \cite{Bartl:1999bc} and we use the same hadronic 
parameters. Moreover, one can also
calculate the SUSY contributions to the 
$\rho$-parameter as given  in \cite{Drees:1990dx}.

\section{Extensions to SLHA}
\label{sec:slhaextensions}

In this section we describe the \spheno\ specific extensions to the
SUSY Les Houches Accord (SLHA) \cite{Skands:2003cj,Allanach:2008qq}.
We start first with extensions to existing blocks and then discuss
new blocks which either control the behaviour of \spheno\ or contain
additional model parameters for MSSM extensions. Note, that
all additional Yukawa couplings have been implemented in complex
forms and the corresponding information can be passed by
using the corresponding blocks starting with {\tt IM}
 \cite{Allanach:2008qq}. 

\subsection{Extensions of existing blocks}
 
\subsubsection{Block {\tt MINPAR}}
\label{sec:minpar}
In the case that generation mixing is switched on, i.e.\ the entry
6 contains a non-zero value, then independent of this value
 flavour violation is switched on in the (s)lepton as
well as in the (s)quark sector.

In case of extending the model by a minimal $SU(5)$ as used in
\cite{Deppisch:2007xu} this block
gets extended by the following entries
\begin{itemize}
\item[7:] $SO(10)$ scale where the universal soft SUSY breaking 
  parameters are defined.
\item[8:] extra  $D$-terms due to the breaking of $SO(10)$ to $SU(5)$
\item[9:] $\lambda$-coupling of the Higgs $24$-plet to the $\bar{5}_H$  
\item[10:] $\lambda'$-coupling of the Higgs $24$-plet to the $5_H$  
\end{itemize}

\subsubsection{Block {\tt MODSEL}}
\label{sec:modsel}

Four switches have been added to flag 3 (particle content)
\begin{itemize}
\item[2:] includes the particle content of a minimal $SU(5)$ model between
   $M_{\rm GUT}$ and a user chosen $SO(10)$ scale, where the SUSY boundary
   conditions are set. The details of this model are described
   in \cite{Deppisch:2007xu}. In this case the mass parameters of 
   the right handed neutrinos
   are stored in the block {\tt MNURIN} (section \ref{sec:mnur}) 
   and the corresponding neutrino
   Yukawa couplings can be stored in the block {\tt YNU0IN}
   (section \ref{sec:Ynu0}). The data is understood to be defined at the 
   GUT-scale. The additional $SU(5)$ parameters as well as the
   $SO(10)$ scale are specified as extensions of the block {\tt MINPAR},
   see section \ref{sec:minpar}
\item[3:] includes three right-handed (s)neutrinos with a common mass for
   all three neutrinos. The neutrino Yukawa couplings $Y_\nu$ can be
  specified at the GUT-scale, see section~\ref{sec:Ynu0}, and the  mass of the
  right-neutrinos at their proper scale, see section~\ref{sec:mnur}.
\item[4:] includes three right-handed (s)neutrinos which are included at their
   proper mass scale. The neutrino Yukawa couplings $Y_\nu$ can be
  specified at the GUT-scale, see section~\ref{sec:Ynu0}, and the  masses of the
  right-neutrinos at their proper scale, see section~\ref{sec:mnur}.
\item[5:] includes a Higgs triplet (15-plet) to realize the seesaw II where the formulas
  of \cite{Rossi:2002zb} including
the corrections presented in \cite{Borzumati:2009hu}  and the
2-loop contributions to the RGEs of the gauge couplings and gaugino
mass parameters have been implemented. 
The additional model data are specified in
   the  blocks \texttt{Higgs3IN} and {\tt YT0}, see sections
\ref{sec:blockhiggs3} and \ref{sec:blockyt0},respectively.
\item[10:] includes three Higgs 24-plets to realize the seesaw 
type III where 
the complete 2-loop RGEs as given in \cite{Esteves:2010ff}
are used.   The additional model data are specified in
   the  blocks \texttt{MWMIN} and {\tt YB3IN}, see sections
 \ref{sec:blockmwmin} and \ref{sec:blockybin}, respectively.
\item[11:] includes a Higgs 15-plet to realize the seesaw II where 
the complete 2-loop RGEs as given in \cite{Esteves:2010ff}
are used.   The additional model data are specified in
the  blocks \texttt{Higgs3IN} and {\tt YTIN}, see sections
 \ref{sec:blockhiggs3} and \ref{sec:blockytin}, respectively.
\end{itemize}

\subsection{New input blocks}
In the output the blocks will be given without the ending IN.

\subsubsection{Block {\tt Higgs3IN}}
\label{sec:blockhiggs3}

Used to specifiy the model of the seesaw model type II. The
data are given in the format
    \begin{verbatim}
(2x,i3,2x,1p,e16.8,0p,2x,'# ',a)
\end{verbatim} and the entries correspond to
 \begin{itemize}
\item[1:] mass of the Higgs triplet (15-plet)
\item[2:] real part of $\lambda_1$
\item[3] imaginary part of $\lambda_1$
\item[4:] real part of $\lambda_2$
\item[5] imaginary part of $\lambda_2$
\item[6:] if 0 (1) use RGEs for the triplet (15-plet) case
\end{itemize}

\subsubsection{Block {\tt MNURIN}}
\label{sec:mnur}

In this block one can specify the masses of the right-handed neutrinos
within the seesaw I model. The masses $m_{Ri}$ are specified in the \fort format
 \begin{verbatim}
(1x,i3,3x,1p,e16.8,3x,'#',a).
\end{verbatim}  

 \subsubsection{Block {\tt MWMIN}} 
 \label{sec:blockmwmin}
Here one can specify the  mass matrix of the 24-plets $M_{Wij}$ 
at $M_{\rm GUT}$ for the seesaw type III model using the 
formulas of \cite{Esteves:2010ff}, 
 where the data are given in the \fort format 
 \begin{verbatim}
(1x,2i3,3x,1p,e16.8,3x,'#',a)
\end{verbatim}  
 where the first two integers in the format correspond to $i$ and $j$ and the double
precision number to the mass parameter.

\subsubsection{Block {\tt NeutrinoBoundsIn}}
\label{sec:neutrinoboundsin}

\begin{table}
\label{tab:neutrinodata}
\caption{Default values for fitting R-parity violating parameters 
if the entries in block {\tt NeutrinoBoundsIn}
are not specified. The values are taken from \cite{Schwetz:2011qt} and correspond to the
1 $\sigma$ range but for $|U_{e3,max}|^2$ which is 90\% CL.
}
\begin{center}
\begin{tabular}{|cc|cc|cc|}
\hline 
$\tan^2 \theta_{atm,min}$ & 0.8182 &
$\tan^2 \theta_{sol,min}$ & 0.4286 &
$U^2_{e3,min}$ & 0 \\
$\tan^2 \theta_{atm,max}$ & 1.3256 &
$\tan^2 \theta_{sol,max}$ & 0.4970 &
$|U_{e3,max}|^2$ & 0.035 \\ \hline 
$\Delta m^2_{atm,min}$ & $2.36 \cdot 10^{-3}$ eV$^2$ &
$\Delta m^2_{sol,min}$ & $7.46 \cdot 10^{-5}$ eV$^2$ & & \\
$\Delta m^2_{atm,max}$ & $2.54 \cdot 10^{-3}$ eV$^2$ &
$\Delta m^2_{sol,max}$ & $7.83 \cdot 10^{-5}$ eV$^2$ & & \\ \hline
\end{tabular}
\end{center}
\end{table}
One can use \spheno\ to obtain R-parity violating parameters consistent
with neutrino data. The corresponding default values are given in 
table \ref{tab:neutrinodata}.
This block can be used to modify them.
 The 
\fort format is
\begin{itemize}
\item[] \begin{verbatim}(1x,i2,3x,1p,e16.8,0p,3x,’#’,1x,a)} \end{verbatim}
\end{itemize}
 and the entries correspond to
\begin{itemize}
\item[1:] $\Delta m^2_{atm,min}$ \dots lower bound on the athmospheric mass difference
\item[2:] $\Delta m^2_{atm,max}$ \dots upper bound on the athmospheric mass difference
\item[3:] $\tan^2 \theta_{atm,min}$ \dots lower bound on the $\tan$ squared of the
         athmospheric mixing angle
\item[4:] $\tan^2 \theta_{atm,max}$ \dots upper bound on the $\tan$ squared of the
         athmospheric mixing angle
\item[5:] $\Delta m^2_{sol,min}$ \dots lower bound on the solar mass difference
\item[6:] $\Delta m^2_{sol,max}$ \dots upper bound on the solar mass difference
\item[7:] $\tan^2 \theta_{sol,min}$ \dots lower bound on the $\tan$ squared of the
         solar mixing angle
\item[8:] $\tan^2 \theta_{sol,max}$ \dots upper bound on the $\tan$ squared of the
         solar mixing angle
\item[9:] $U^2_{e3,min}$ \dots lower bound on the mixing element $U_{e3}$ squared
          (reactor angle)
\item[10:] $U^2_{e3,max}$ \dots upper bound on the mixing element $U_{e3}$ squared
\end{itemize}

\subsubsection{Block {\tt SPhenoInput}}
\label{sec:sphenoinput}
This block sets the \spheno\ specific flags. The 
FORTRAN format is
\begin{itemize}
\item[] \begin{verbatim}(1x,i2,3x,1p,e16.8,0p,3x,’#’,1x,a)} \end{verbatim}
\end{itemize}
 and the entries correspond to
\begin{itemize}
\item[1:] sets the error level
\item[2:] if 1 the the SPA conventions \cite{AguilarSaavedra:2005pw}
are used
\item[3:] takes a spectrum which is given by an external program
\item[4:] introduces an extension of the SLHA output: in the case of
 flavour violation, flavour ordered states are used instead of
mass ordered states.
\item[11:] if 1 then the branching ratios of the SUSY and Higgs particles
are calculated, if 0 then this calculation is omitted.
\item[12:] sets minimum value for a branching ratios, so that it appears in the output
\item[21:] if 1 then the cross sections of SUSY and Higgs particles in $e^+ e^-$ 
annihilation
are calculated, if 0 then this calculation is omitted.
\item[22:] sets the center of mass energy $E_{cms}$, can be repeated up to 100 times
\item[23:] sets the electron polarisation $P_m$, can be repeated up to 100 times
\item[24:] sets the positron polarisation $P_p$, can be repeated up to 100 times
\item[25:] whether to use initial state radation in the calculation of the cross
sections
\item[26:] sets minimum value for a cross section, so that it appears in the output
\item[31:] sets the value of $M_{\rm GUT}$, otherwise $M_{\rm GUT}$
is determined by the condition $g_1=g_2$
\item[32:] sets strict unification, i.e.\  $g_1=g_2=g_3$

\item[34:] sets the relative precision with which the masses are
calculated, default is $10^{-6}$ 
\item[35:] sets the maximal number of iterations in the calculation
 of the masses, default  is 40
\item[36:] whether to write out debug information for the loop calculations
\item[38:] this entry sets the loop order of the RGEs: either 1 or 2, default
 is 2, i.e.\ using  2-loop RGEs 
\item[41:] sets the width of the Z-boson $\Gamma_Z$,
default is 2.49 GeV
\item[42:] sets the width of the W-boson $\Gamma_W$,
default is 2.06 GeV
\item[80:] if not set 0  the program exists with a non-zero value  if
           a problem has occured
\item[90:] if 1 add R-parity to a high scale spectrum calculated either
       from mSUGRA, GMSB or AMSB boundary conditions
\item[91:] if 1 than bilinear parameters are calculated such that neutrino data
           are fitted in the experimenatal allowed range (the range can be changed
           using the Block {\tt NeutrinoBoundsIn}, see section \ref{sec:neutrinoboundsin})
\item[92:] if 1 gives in case of R-parity violation only the $4\times 4$ MSSM
           part of the neutrino/neutralino mixing matrix $N$ 
           and the correspondingly
           the $2 \times 2$ parts of the charged lepton/chargino mixing 
           matrices $U$ and $V$ as well as the block for the stau mixing.
            This is in particular useful in case one uses the program
           {\tt Prospino} \cite{Beenakker:1996ed} or older
           versions of the program {\tt Phythia} \cite{Sjostrand:2006za}.
\end{itemize}

 \subsubsection{Block {\tt YB3IN}}
 \label{sec:blockybin}
Here one can specify the   neutrino Yukawa $Y^{III}_{ij}$ coupling 
at $M_{\rm GUT}$ for the seesaw type III model using the 
formulas of \cite{Esteves:2010ff}, 
 where the data are given in the \fort format 
 \begin{verbatim}
(1x,2i3,3x,1p,e16.8,3x,'#',a)
\end{verbatim}  
 where the first two integers in the format correspond to $i$ and $j$ and the double
precision number to Yukawa coupling.

\subsubsection{Block {\tt YNU0IN}}
\label{sec:Ynu0}

This block specifies the neutrino Yukawa couplings $Y_\nu$ at the GUT scale
and the corresponding superpotential term is given by
$W = Y_{\nu,ij} \hat{\nu}^C_{i} \hat{L}_j \hat{H}_u$. 
It is assumed that the right-handed neutrinos are in the mass eigenbasis.
The real parts are specified in the block \verb+YNU0IN+ with the FORTRAN format
\begin{verbatim}
(1x,i2,1x,i2,3x,1p,e16.8,0p,3x,'#',1x,a)
\end{verbatim}
and the imaginary parts in the block \verb+IMYNU0IN+ 
with the same FORTRAN input.

 \subsubsection{Block {\tt YT0IN}} 
 \label{sec:blockyt0}
Here one can specify the   neutrino Yukawa $Y^T_{ij}$ coupling 
at $M_{\rm GUT}$ for the seesaw type II model using the 
formulas of \cite{Rossi:2002zb,Borzumati:2009hu}.
The data is given in the \fort format 
 \begin{verbatim}
(1x,2i3,3x,1p,e16.8,3x,'#',a)
\end{verbatim}  
 where the first two integers in the format correspond to $i$ and $j$ and the double
precision number to Yukawa coupling.

 \subsubsection{Block {\tt YTIN}} 
 \label{sec:blockytin}
Here one can specify the   neutrino Yukawa $Y^T_{ij}$ coupling 
at $M_{\rm GUT}$ for the seesaw type II model using the 
formulas of \cite{Esteves:2010ff}, 
 where the data is given in the \fort format 
 \begin{verbatim}
(1x,2i3,3x,1p,e16.8,3x,'#',a)
\end{verbatim}  
 where the first two integers in the format correspond to $i$ and $j$ and the double
precision number to Yukawa coupling.

\subsection{New output block}
\subsubsection{Block {\tt SPhenoLowEnergy}}

In this block the calculated values of the low energy observables are 
given:
\begin{itemize}
\item[1] $BR(b \to s \gamma)$
\item[2] $BR(b \to s \mu^+ \mu^-)$
\item[3] $BR(b \to s \sum_i \nu_i \nu_i)$
\item[4] $BR(B^0_d \to \mu^+ \mu^-)$
\item[5] $BR(B^0_s \to \mu^+ \mu^-)$
\item[6] $BR(B_u \to \tau^+ \nu)$
\item[7] $BR(B_u \to \tau^+ \nu)/BR(B_u \to \tau^+ \nu)_{SM}$
\item[8] $\Delta(M_{B_s^0})$ [in ps$^{-1}$]
\item[9] $\Delta(M_{B_d^0})$ [in ps$^{-1}$]
\item[20] SUSY contribution to the anomalous magnetic moment of the electron $\Delta(\frac{g-2}{2})_e$
\item[21] SUSY contribution to the anomalous magnetic moment of the muon $\Delta(\frac{g-2}{2})_\mu$
\item[22] SUSY contribution to the anomalous magnetic moment of the tau $\Delta(\frac{g-2}{2})_\tau$
\item[23] electric dipole moment of the electron $d_e$
\item[24] electric dipole moment of the muon $d_\mu$
\item[25] electric dipole moment of the tau $d_\tau$
\item[26] $BR(\mu\to e \gamma)$
\item[27] $BR(\tau\to e \gamma)$
\item[28] $BR(\tau\to \mu \gamma)$
\item[29] $BR(\mu^+\to e^+ e^+ e^-)$
\item[30] $BR(\tau^+\to e^+ e^+ e^-)$
\item[31] $BR(\tau^+\to \mu^+ \mu^+ \mu^-)$
\item[39] SUSY contribution to the $\rho$-parameter
\item[40] $BR(Z^0 \to e^\pm \mu^\mp)$
\item[41] $BR(Z^0 \to e^\pm \tau^\mp)$
\item[42] $BR(Z^0 \to \mu^\pm \tau^\mp)$

\end{itemize}

\section{Installation and implementing new models}

\subsection{Installation}

\spheno\ can be downloaded from 
\begin{center}
\verb+ http://projects.hepforge.org/spheno/+
\end{center}
where the latest tar-ball {\tt SPheno3.x.y.tar.gz} can found
as well as older versions.
Unpacking will create the directory {\tt  SPheno3.x.y} where
{\tt x} and {\tt y} are integers corresponding to the sub-version.
This directory will contain the following subdirectories:
\begin{itemize}
\item {\tt bin}: here the executable \spheno\ will be stored
\item {\tt doc}: contains the \spheno\ documentations
\item {\tt include}: here all the mod-files are stored
\item {\tt input}: contains input example files
\item {\tt lib}: here the library  {\tt libSPheno.a} will be stored
\item {\tt output}: contains the output files corresponding
to the examples stored in  {\tt input}
\item {\tt  src}: contains the source code
\end{itemize}
The directory {\tt  SPheno3.x.y} contains a Makefile which can
be used to compile \spheno. The default compiler is Intels ifort, but
by typing {\tt make F90=compiler} on the console one can use
a different compiler where {\tt compiler} has to replaced by the compiler's
name. The following compilers have been added
NAG nagfor, Lahey lf95 and g95. 

It is well known that compilation of the module {\tt RGEs.F90}
can be time consuming due to the length of the 2-loop RGEs for
the seesaw models of type II and type III. For this reason they
are not compiled by default. If the corresponding RGEs should be
included then the line \\ 
\centerline{\tt PreDef = -DGENERATIONMIXING  -DONLYDOUBLE} 
should be replaced by \\
\centerline {\tt PreDef = -DGENERATIONMIXING  -DONLYDOUBLE -DSEESAWIII}
i.e.\ add {\tt -DSEESAWIII}.

In the case that one want to have quadrupole precision in various
parts of the code instead of double precision, one has to take out
the {\tt  -DONLYDOUBLE} in the line mentioned above. Note that this can
substantially slow down \spheno. Moreover, not all parts are yet
implemented with quadrupole precision. The main focus has
been on the loop functions as well as on mixing between 
neutralinos and neutrinos in case of R-parity violation.

\subsection{Implementing new models}
New models can easily implemented using the \sarah package
\cite{Staub:2008uz,Staub:2010jh}. For this purpose one has
to put the code generated by \sarah in
a new directory within the directory {\tt  SPheno3.x.y} and run
the corresponding Makefile. An additional  executable will
be stored in the directory {\tt bin}.

\section{Input and output}

Starting with version \spheno\ 3.1 there are two main differences
with respect to the input and output
\begin{enumerate}
\item \spheno\ accepts only the SLHA input format as specified
  and all the
 output is given in this format. In section \ref{sec:slhaextensions}
 we have described the extensions to control program specific features
 as well as model extensions.
 The orginal \spheno\ input
 using the files {\tt HighScale.in}, {\tt StandardModel.in} and
 {\tt Control.in}  as well as the output in the file
 {\tt SPheno.out} have been disabled. Detailed error messages and warnings
 will also be written to the file {\tt Messages.out}.
\item One can provide input name and output name as command line
 options where the first (second) name, if present, is interpreted
 as input (output) filename, e.g.
 
 \centerline{ \spheno\ \tt InName OutName}

 takes {\tt InName} for the file containing the input and will write
 the output to the file {\tt OutName}. In case that the file {\tt InName} 
 is not found \spheno\ will look for a file called {\tt LesHouches.in}
 as default. The default name for the output is {\tt SPheno.spc}.
 The length of the names {\tt InName} and {\tt OutName} must not
 exceed 60 characters.
\end{enumerate}

\section{Conclusions and comments}

\spheno\ is  constantly developing, in particular in view
of implementing additional models and low energy observables.
In addition it is planed
\begin{itemize}
\item
 to implement the missing pieces
of the SLHA conventions as listed in appendix \ref{sec:unsupportedslha}.
\item mixing between $A_0$ and $H_0$ in case of CP phases 
\item low energy observables for the case of R-parity violation
\end{itemize}
In section \ref{sec:lowenergy} several hadronic parameters for
the calculation of low energy observables are hard-coded in the
program. It is planned to construct routines to allow user defined
changes in the future.

\section*{Acknowledgements}

W.P.\ thanks M.~Herrero for providing routines for the calculation
of the flavour violating three body decays of leptons. 
The authors thank M.~Hirsch for many valuable comments on the program
and its handling. This work  has been
supported by DFG, project number PO-1337/1-1.

\begin{appendix}

\section{Default SM values}

The following default values will be used if not given in the file \verb+LesHouches.in+.
\begin{itemize}
\item CKM-matrix, Wolfenstein parameters: $\lambda = 0.2265$, $A=0.807$,
$\rho = 0.141$, $\eta = 0.343$
\item gauge sector: $1/\alpha_{em}(0)= 137.0359895$, $m_Z = 91.187$~GeV, 
$G_F=1.16637\cdot 10^{-5}$GeV$^{-2}$, \\
$\alpha_s^{\overline{MS}}(m_Z) = 0.1184$
\item lepton masses: $m_e= 510.99891$~keV, $m_\mu=105.658$~MeV, $m_\tau=1.7768$~GeV
\item quark masses: $m_u(2$ GeV$)= 3$~MeV, $m_d(2$ GeV$)= 5$~MeV,
 $m_s(2$ GeV$)= 105$~MeV,
$m_c(m_c)=1.27$~GeV, $m_b(m_b)=4.2$~GeV, $m_t=171.3$~GeV;
 the top mass is interpreted
as on-shell mass
\end{itemize}

\section{Unsupported SLHA features}
\label{sec:unsupportedslha}
Here we list the features of the SLHA conventions 
\cite{Skands:2003cj,Allanach:2008qq} which are not yet
supported:
\begin{itemize}
\item In {\tt Block MODSEL} the following entries are currently ignored:
 \begin{enumerate}
  \item[11:] the possiblity to give the parameters at $n$ equidistant values of
             $Q$ between $m_Z$ and $Q_{EWSB}$
  \item[21:] to give the mass parameters at the pole mass of each individual particle
 \end{enumerate}

\item In {\tt Block EXTPAR} the following entries are currently ignored:
 \begin{enumerate}
  \item[27:] pole mass of the charged Higgs boson
  \item[51:] (GMSB only) $U(1)_Y$ messenger index
  \item[52:] (GMSB only) $SU(2)_L$ messenger index
  \item[53:] (GMSB only) $SU(3)_C$ messenger index
 \end{enumerate}
\item the {\tt Block QEXTPAR}
\item the {\tt Block RVLAMLLEIN}
\item the {\tt Block RVLAMLQDIN}
\item the {\tt Block RVLAMUDDIN}
\item the {\tt Block RVTLLEIN}
\item the {\tt Block RVTLQDIN}
\item the {\tt Block RVTUDDIN}
\item the {\tt Block RVDIN}
\item the {\tt Block RVM2LH1IN}
\end{itemize}
These features will be implemented within the next updates.

\section{Error messages and warnings, interpretation of the variable kont}

Here we describe how to interpret the values of the variable {\tt kont} which
is used in the error system of \spheno. The corresponding warnings and
error messages are also given in the file 'Messages.out' if the error
level is set to the appropriate value.

\subsection{Module Mathematics}

\begin{itemize}
\item[-1:] step size gets to small in routine {\tt ODEint}
\item[-2:] maximal value $> 10^{36}$ {\tt ODEint}
\item[-3:] too many steps are required in routine  {\tt ODEint}
\item[-4:] boundary conditions cannot fullfilled in routine {\tt ODEintB}
\item[-5:] maximal value $> 10^{36}$ {\tt ODEintB}
\item[-6:] step size gets too small in routine  {\tt ODEintB}
\item[-7:] too many steps are required in routine   {\tt ODEintB}
\item[-8:] boundary conditions cannot fullfilled in routine {\tt ODEintC}
\item[-9:] maximal value $> 10^{36}$ {\tt ODEintC}
\item[-10:] step size gets too small in routine {\tt ODEintC} 
\item[-11:] too many steps are required in routine  {\tt ODEintC}
\item[-12:] step size gets too small in routine {\tt rkqs}
\item[-13:] the size of the arrays do not match in routine
 {\tt ComplexEigenSystems}
\item[-14:] potential numerical problems in routine {\tt ComplexEigenSystems}
\item[-15:] the size of the arrays do not match in routine {\tt RealEigenSystemse}
\item[-16:] potential numerical problems in routine {\tt RealEigenSystems}
\item[-17:] the size of the arrays do not match in routine {\tt tqli}
\item[-18:] too many iterations in routine {\tt tqli}
\item[-19:] too high accuracy required in routine {\tt Dgauss}
\item[-20:] too high accuracy required in routine {\tt DgaussInt}
\item[-21:] precision problem in routine {\tt Kappa}
\item[-22:] step size gets too small in routine {\tt IntRomb} 
\item[-23:] too many steps are required in routine  {\tt IntRomb}
\item[-24:] singular matrix in routine {\tt GaussJ}
\item[-25:] inversion failed in routine {\tt InvMat3}
\item[-26:] singular matrix in routine {\tt GaussJ}
\end{itemize}

\subsection{Module StandardModel}

\begin{itemize} 
\item[-101:] routine {\tt CalculateRunningMasses}: $Q_{low} > m_b(m_b)$
\item[-102:] routine {\tt CalculateRunningMasses}: 
Max($Q_{low},m_b(m_b) > Q_{max})$
\end{itemize}

\subsection{Module SusyMasses}

\begin{itemize} 
\item[-201:] negative mass squared in routine {\tt ChargedScalarMassEps1nt}
\item[-202:] negative mass squared in routine {\tt ChargedScalarMassEps3nt}
\item[-204:] $|Y_\tau|^2 < 0$ in routine {\tt CharginoMass3}
\item[-205:] $|Y_\tau|^2 < 0$ in routine {\tt CharginoMass5}
\item[-206:] negative mass squared in routine {\tt PseudoScalarMassEps1nt}
\item[-207:] negative mass squared in routine {\tt PseudoScalarMassEps3nt}
\item[-208:] negative mass squared in routine {\tt PseudoScalarMassMSSMnt}
\item[-210:] negative mass squared in routine {\tt ScalarMassEps1nt}
\item[-211:] negative mass squared in routine {\tt ScalarMassEps3nt}
\item[-212:] negative mass squared in routine {\tt ScalarMassMSSMeff}
\item[-213:] negative mass squared in routine {\tt ScalarMassMSSMnt}
\item[-215:] $m^2_{S^0_1} < 0$ in routine {\tt ScalarMassMSSMeff}
\item[-216:] $m^2_{P^0_1} < 0$ in routine {\tt ScalarMassMSSMeff}
\item[-217:] $m^2_{S^+} < 0$ in routine {\tt ScalarMassMSSMeff}
\item[-220:] negative mass squared in routine {\tt SfermionMass1Eps1}
\item[-221:] negative mass squared in routine {\tt SfermionMass1Eps3}
\item[-222:] negative mass squared in routine {\tt SfermionMass1MSSM}
\item[-223:] negative mass squared in routine {\tt SfermionMass3MSSM}
\item[-224:] negative mass squared in routine {\tt SquarkMass3Eps}
\item[-225:] $m^2_{\tilde \nu} < 0$ in routine {\tt TreeMassesEps1}
\item[-226:] $m^2_{\tilde \nu} < 0$ in routine {\tt TreeMassesMSSM}
\item[-227:] $m^2_{A^0} < 0$ in routine {\tt TreeMassesMSSM}
\item[-228:] $m^2_{H^+} < 0$ in routine {\tt TreeMassesMSSM}
\item[-229:] $m^2_{\tilde \nu} < 0$ in routine {\tt TreeMassesMSSM2}
\item[-230:] $m^2_{A^0} < 0$ in routine {\tt TreeMassesMSSM2}
\item[-231:] $m^2_{H^+} < 0$ in routine {\tt TreeMassesMSSM2}
\item[-232:] $m^2_{\tilde \nu} < 0$ in routine {\tt TreeMassesMSSM3}
\end{itemize}

\subsection{Module InputOutput}

\begin{itemize} 
\item[-302:] routine {\tt LesHouches\_Input}: unknown entry for Block MODSEL
\item[-303:] routine {\tt LesHouches\_Input}: model must be specified before parameters 
\item[-304:] routine {\tt LesHouches\_Input}: unknown entry for Block MINPAR
\item[-305:] routine {\tt LesHouches\_Input}: model has not been specified completly
\item[-306:] routine {\tt LesHouches\_Input}: a serious error has been part of the input
\item[-307:] routine {\tt LesHouches\_Input}: Higgs sector has not been fully specified
\item[-308:] routine {\tt ReadMatrixC}: indices exceed the given boundaries  
\item[-309:] routine {\tt ReadMatrixR}: indices exceed the given boundaries 
\item[-310:] routine {\tt ReadVectorC}: index exceeds the given boundaries 
\item[-311:] routine {\tt ReadVectorR}: index exceeds the given boundaries 
\item[-312:] routine {\tt ReadMatrixC}: indices exceed the given boundaries 
\end{itemize}

\subsection{Module SugraRuns}

\begin{itemize} 
\item[-401:] routine {\tt BoundaryEW}: negative scalar mass squared as input
\item[-402:] routine {\tt BoundaryEW}: $m_Z^2(m_Z) < 0$
\item[-403:] routine {\tt BoundaryEW}: $\sin^2 \theta_{\overline{DR}} < 0$
\item[-404:] routine {\tt BoundaryEW}: $m_W^2 < 0$
\item[-405:] routine {\tt BoundaryEW}: either $m_{l_DR}/m_l < 0.1$ or $m_{l_DR}/m_l > 10$
\item[-406:] routine {\tt BoundaryEW}: either $m_{d_DR}/m_u < 0.1$ or $m_{d_DR}/m_d > 10$
\item[-407:] routine {\tt BoundaryEW}: either $m_{u_DR}/m_d < 0.1$ or $m_{u_DR}/m_u > 10$
\item[-408:] routine {\tt RunRGE}:  entering non-perturbative regime 
\item[-409:] routine {\tt RunRGE}:  nor $g_1 \ne g_2$ at $M_{\rm GUT}$
 neither any other unification
\item[-410:] routine {\tt RunRGE}:  entering non-perturbative regime
 at $M_{\rm GUT}$
\item[-411:] routine {\tt RunRGE}:  entering non-perturbative regime at $M_{H_3}$
\item[-412:] routine {\tt Sugra}: run did not converge    
\item[-413:] routine {\tt Calculate\_Gi\_Yi}: $m_Z^2(m_Z) < 0$
\item[-414:] routine {\tt Calculate\_Gi\_Yi}: too many iterations to
    calculate $m_b(m_b)$ in the $\overline{MS}$ scheme
\item[-415:] routine {\tt Sugra}: $|\mu|^2 < 0$ at $m_Z$
\end{itemize}

\subsection{Module LoopMasses}
\begin{itemize}
\item[-501] negative mass squared in routine {\tt SleptonMass\_1L}
\item[-502] $p^2$ iteration did not converge in routine {\tt SleptonMass\_1L}
\item[-503] negative mass squared in routine {\tt SneutrinoMass\_1L}
\item[-504] $p^2$ iteration did not converge in routine {\tt SneutrinoMass\_1L}
\item[-505] negative mass squared in routine {\tt SquarkMass\_1L}
\item[-506] $p^2$ iteration did not converge in routine {\tt SquarkMass\_1L}
\item[-507] $m_{h^0}^2 < 0$  in routine {\tt LoopMassesMSSM}
\item[-508] $m_{A^0}^2 < 0$  in routine {\tt LoopMassesMSSM}
\item[-509] $m_{H^+}^2 < 0$  in routine {\tt LoopMassesMSSM}
\item[-510] $|\mu|^2 > 10^{20}$  in routine {\tt LoopMassesMSSM}
\item[-511] $|\mu|^2 < 0$  in routine {\tt LoopMassesMSSM}
\item[-512] $m^2_Z(m_Z)^2 < 0$  in routine {\tt LoopMassesMSSM}
\item[-513] $m_{h^0}^2 < 0$  in routine {\tt LoopMassesMSSM\_2}
\item[-514] $m_{A^0}^2 < 0$  in routine {\tt LoopMassesMSSM\_2}
\item[-515] $m_{H^+}^2 < 0$  in routine {\tt LoopMassesMSSM\_2}
\item[-516] $|\mu|^2 > 10^{20}$  in routine {\tt LoopMassesMSSM\_2}
\item[-517] $|\mu|^2 < 0$  in routine {\tt LoopMassesMSSM\_2}
\item[-518] $m^2_Z(m_Z)^2 < 0$  in routine {\tt LoopMassesMSSM\_2}
\item[-519] $m_{h^0}^2 < 0$  in routine {\tt LoopMassesMSSM\_3}
\item[-520] $m_{A^0}^2 < 0$  in routine {\tt LoopMassesMSSM\_3}
\item[-521] $m_{H^+}^2 < 0$  in routine {\tt LoopMassesMSSM\_3}
\item[-522] $|\mu|^2 > 10^{20}$  in routine {\tt LoopMassesMSSM\_3}
\item[-523] $|\mu|^2 < 0$  in routine {\tt LoopMassesMSSM\_3}
\item[-524] $m^2_Z(m_Z)^2 < 0$  in routine {\tt LoopMassesMSSM\_3}
\end{itemize}

\subsection{Module TwoLoopHiggsMass}

\begin{itemize} 
\item[-601:] routine {\tt PiPseudoScalar2}: $m^2_{\tilde t} < 0$
\item[-602:] routine {\tt PiPseudoScalar2}: $m^2_{\tilde b} < 0$
\item[-603:] routine {\tt PiPseudoScalar2}: $m^2_{\tilde \tau} < 0$
\item[-604:] routine {\tt PiScalar2}: $m^2_{\tilde t} < 0$
\item[-605:] routine {\tt PiScalar2}: $m^2_{\tilde b} < 0$
\item[-606:] routine {\tt PiScalar2}: $m^2_{\tilde \tau} < 0$
\item[-607:] routine {\tt Two\_Loop\_Tadpoles}: $m^2_{\tilde t} < 0$
\item[-608:] routine {\tt Two\_Loop\_Tadpoles}: $m^2_{\tilde b} < 0$
\item[-609:] routine {\tt Two\_Loop\_Tadpoles}: $m^2_{\tilde \tau} < 0$
\end{itemize}

\subsection{Module MathematicsQP}

\begin{itemize}
\item[-1001:] the size of the arrays do not match in routine
 {\tt ComplexEigenSystems\_DP}
\item[-1002:] potential numerical problems in routine {\tt ComplexEigenSystems\_DP}
\item[-1003:] the size of the arrays do not match in routine
 {\tt ComplexEigenSystems\_QP}
\item[-1004:] potential numerical problems in routine {\tt ComplexEigenSystems\_QP}
\item[-1005:] the size of the arrays do not match in routine {\tt RealEigenSystems\_DP}
\item[-1006:] potential numerical problems in routine {\tt RealEigenSystems\_DP}
\item[-1007:] the size of the arrays do not match in routine {\tt RealEigenSystems\_QP}
\item[-1008:] the size of the arrays do not match in routine {\tt Tqli\_QP}
\item[-1009:] too many iterations in routine {\tt Tqli\_QP}
\item[-1010:] too many iterations in routine {\tt Tql2\_QP}
\end{itemize}

\section{Loop corrections}

Here we list the improvements which have been implemented in
\spheno\ with respect to ref.~\cite{Pierce:1996zz}:
\begin{itemize}
\item in the 1-loop corrections to the gluino mass we use for the
gluon contribution
\begin{equation}
\Delta(\Sigma_{\tilde g}) = - \frac{3 g^2_3}{8 \pi^2}
\left( B_1(p^2, m^2_{\tilde g,T},0) - 2  B_1(p^2, m^2_{\tilde g,T},0) \right)
\end{equation}
where $m_{\tilde g,T}$ is the tree level gluino mass and 
which reduces for $p^2 = m^2_{\tilde g,T}$ to the formula
\begin{equation}
\Delta(\Sigma_{\tilde g}) = - \frac{g^2_3}{16 \pi^2}
\left( 15 + 9 \log  \left( \frac{Q^2}{m^2_{\tilde g,T}} \right) \right)
\end{equation}
of ref.~\cite{Pierce:1996zz}.
\item 
In addition flavour violation has been taking into account and the
corresponding formulas can be found in \cite{Bruhnke:2010rh,Staub:2010ty}.
\end{itemize}

\end{appendix}


\end{document}